\documentclass[showpacs,prl,amsmath,twocolumn,nofootinbib]{revtex4}

\usepackage{float,placeins}
\usepackage{amsmath}
\usepackage{amssymb}
\usepackage{mathtools}
\usepackage{subfigure}
\usepackage{epsfig}
\usepackage{listings}
\usepackage{enumitem}
\usepackage{graphicx}    
\usepackage{graphics}
\usepackage{longtable}
\usepackage{hyperref}
\usepackage{breakurl}
\usepackage{epigraph}
\usepackage{xspace}

\newcommand{\elabel}[1]{\label{eq:#1}}
\newcommand{\eref}[1]{Eq.~(\ref{eq:#1})}

\newcommand{\Eref}[1]{Equation~(\ref{eq:#1})}

\newcommand{\ie}{{\it i.e.}\xspace}
\newcommand{\eg}{{\it e.g.}\xspace}

\newcommand{\ave}[1]{\left\langle#1 \right\rangle}

\newcommand{\flabel}[1]{\label{fig:#1}}
\newcommand{\fref}[1]{Fig.~\ref{fig:#1}}
\newcommand{\Fref}[1]{Figure ~\ref{fig:#1}}

\newcommand{\be}{\begin{equation}}
\newcommand{\ee}{\end{equation}}

\newcommand{\bi}{\begin{itemize}}
\newcommand{\ei}{\end{itemize}}

\newcommand{\Dt}{\Delta t}
\newcommand{\etau}{\tau^\text{eqm}}
\newcommand{\wtau}{\widetilde{\tau}}
\newcommand{\xN}{\ave{x}_N}
\newcommand{\Sdata}{S^{\text{data}}}
\newcommand{\Smodel}{S^{\text{model}}}

\begin{document}
\title{Far from equilibrium: Wealth reallocation in the United States}
\author{Yonatan Berman}
\email{yonatanb@post.tau.ac.il}
\affiliation{School of Physics and Astronomy, Tel-Aviv University, Tel-Aviv, Israel}
\author{Ole Peters}
\email{o.peters@lml.org.uk}
\affiliation{London Mathematical Laboratory, 14 Buckingham Street, London WC2N 6DF, UK}
\affiliation{Santa Fe Institute, 1399 Hyde Park Road, Santa Fe, NM 87501, USA}
\author{Alexander Adamou}
\email{a.adamou@lml.org.uk}
\affiliation{London Mathematical Laboratory, 14 Buckingham Street, London WC2N 6DF, UK}

\date{\today}

\begin{abstract} 
Studies of wealth inequality often assume that an observed wealth distribution reflects a system in equilibrium. This constraint is rarely tested empirically. We introduce a simple model that allows equilibrium but does not assume it. To geometric Brownian motion (GBM) we add reallocation: all individuals contribute in proportion to their wealth and receive equal shares of the amount collected. We fit the reallocation rate parameter required for the model to reproduce observed wealth inequality in the United States from 1917 to 2012. We find that this rate was positive until the 1980s, after which it became negative and of increasing magnitude. With negative reallocation, the system cannot equilibrate. Even with the positive reallocation rates observed, equilibration is too slow to be practically relevant. Therefore, studies which assume equilibrium must be treated skeptically. By design they are unable to detect the dramatic conditions found here when data are analysed without this constraint.
\end{abstract}


\maketitle



\section{The model -- reallocating GBM}
Our model of personal wealth is geometric Brownian motion (GBM) enhanced with a simple reallocation mechanism. The wealth of the $i^\mathrm{th}$ individual as a function of time, $x_i\left(t\right)$, in a population of $N$ individuals follows the stochastic differential equation,

\be
dx_i = \underbrace{x_i \left(\mu dt + \sigma dW_i\right)}_\mathrm{growth} - \underbrace{ x_i \tau dt + \xN \tau dt}_\mathrm{reallocation}.
\elabel{dx2}
\ee

$dx_i$ is the change in wealth over the time period, $dt$. $dW_i$ is the increment in a Wiener process, which is normally distributed with mean zero and variance $dt$. We refer to the parameters $\mu$ as the drift, $\sigma$ as the volatility, and $\tau$ as the reallocation rate. Angled brackets with subscript $N$ denote the sample mean, \ie $\xN=\frac{1}{N}\sum_{i=1}^N x_i$.

The term labelled `growth' in \eref{dx2} is the wealth increment in a GBM. It is a random proportion of $x_i$, consisting of a certain part, $x_i \mu dt$, and an uncertain part, $x_i \sigma dW_i$. On its own this term generates noisy exponential growth.

The `reallocation' term works as follows: each individual contributes $x_i \tau dt$, an amount proportional to its wealth, and receives $\xN \tau dt$, an equal share of the sum of all contributions. For $\tau>0$, low-wealth individuals with $x_i<\xN$ are net recipients under this mechanism, while high-wealth individuals with $x_i>\xN$ are net contributors. For $\tau<0$, the roles are reversed.

We call this model reallocating geometric Brownian motion (RGBM) to distinguish it from GBM, to which it reduces when $\tau=0$. \Eref{dx2} may also be written as

\be
dx_i = x_i \left(\mu dt + \sigma dW_i\right) - \tau \left( x_i - \xN \right)dt,
\elabel{dx1}
\ee

which shows that RGBM can be viewed as GBM with reversion to the sample mean at rate $\left|\tau\right|$ if $\tau$ is positive, and repulsion from the sample mean at rate $\left|\tau\right|$ if $\tau$ is negative. \Fref{trajectories} depicts typical individual trajectories produced by the RGBM model with positive, zero, and negative reallocation (respectively $\tau>0, ~\tau=0, ~\tau<0$).

\begin{figure*}[!htb]
\centering
\includegraphics[width=0.9\textwidth] {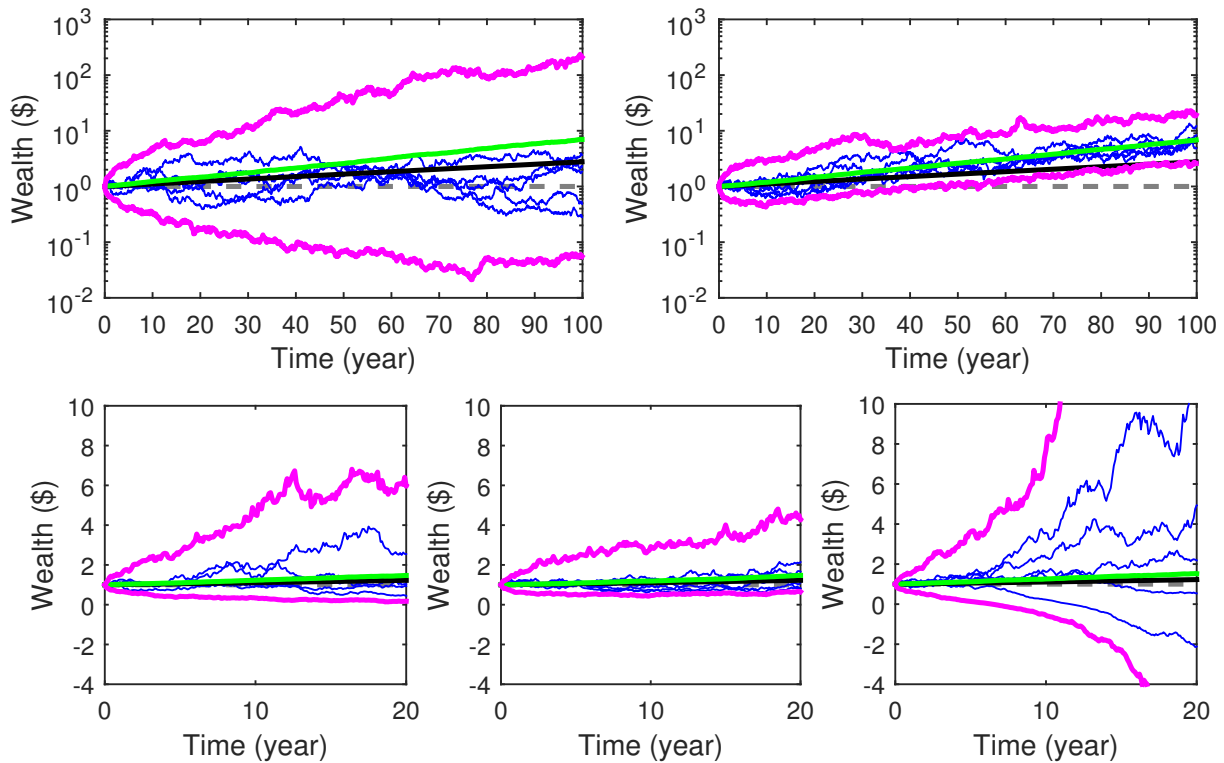}
\caption{Simulations of RGBM with $N=1000$, $\mu=0.021 \text{ year}^{-1}$, $\sigma=0.14\text{ year}^{-1/2}$, $x_i\left(0\right)=\$1$. Effect of reallocation parameter $\tau$. \underline{Top left:} $\tau=0 \text{ year}^{-1}$, inequality grows indefinitely. Magenta lines denote the largest and smallest wealths. The distribution of individual wealth, on logarithmic scales, is symmetric about the median $\exp\left[\left(\mu-\sigma^2/2\right)t\right]$ (black) rather than the sample mean (green). Blue lines illustrate five randomly chosen wealth trajectories. \underline{Top right:} $\tau=0.1\text{ year}^{-1}$, inequality grows initially but is limited by reallocation. The distribution of individual wealth is confined around the sample mean (green), $\xN\approx \exp\left(\mu t\right)$. \underline{Bottom left:} same as top left, zoomed in, and on a linear wealth scale. \underline{Bottom middle:} same as top right, zoomed in, and on a linear wealth scale. \underline{Bottom right:} $\tau=-0.1\text{ year}^{-1}$, the system is in a qualitatively different regime in which wealth can become negative.
\flabel{trajectories}}
\end{figure*}

\subsection{Regimes of RGBM}
The RGBM model produces qualitatively different behaviour for different model parameters, population sizes, and timescales. There are three important regimes characterised by the value of $\tau$. In the following discussion, as in \fref{trajectories}, we let all individuals start with wealth of one dollar, $x_i(0)=\$1$.

\bi
\item[$\tau=0$]
Without reallocation the model is GBM, whose properties are well known (see, for example, \cite{PetersKlein2013}). Wealth follows a lognormal distribution which broadens indefinitely over time. There is no stationary non-zero distribution to which it converges. In relative terms (\ie on logarithmic wealth scales) the wealth distribution is symmetric about the median, $\exp\left[\left(\mu-\sigma^2/2\right)t\right]$.

GBM exhibits an extreme form of a phenomenon known as wealth condensation \cite{BouchaudMezard2000}: over time a measure-zero proportion of the population ends up with a measure-one proportion of wealth. In practical terms, this corresponds to one person holding all the wealth, \ie perfect inequality.

\item[$\tau>0$]
RGBM becomes equivalent to the model of cooperation in \cite{PetersAdamou2015a} as $\tau\to\infty$. In this limit, all individuals have equal wealth which is well approximated by $\exp\left[\left(\mu-\sigma^2/2N\right)t\right]$. For $N$ and $\sigma$ appropriate to human populations, this is very close to $\exp\left(\mu t\right)$.

For finite $\tau$, wealths disperse. However, their distribution remains confined around the sample mean, $\xN$, to which they are connected by the reallocation mechanism. For realistic model parameters, populations and timescales, $\xN \approx \exp\left(\mu t\right)$. The wealth distribution broadens as $\tau\to0^+$ to resemble that of GBM. 

The wealth distribution for RGBM with $\tau>0$ is stationary and non-vanishing, with some diverging moments. As $\tau$ increases, successively higher moments become finite as the distribution becomes more closely confined around the sample mean.

\item[$\tau<0$]
For negative reallocation the dynamics of RGBM are qualitatively different. While $\tau\geq0$ guarantees positive wealth for everyone, with $\tau<0$ the reallocation mechanism pushes individuals with $x_i<\ave{x}_N$ into negative wealth. The creation of large wealths is fuelled not only by multiplicative growth but also by direct transfers from the poor, whose wealths become increasingly negative. No stationary wealth distribution exists. In practical terms this corresponds to creditors and debtors whose wealths diverge exponentially away from the sample mean. 
\ei

\subsection{Interpretation of $\tau$}
In the GBM model (\ie $\tau=0$), some individuals in a population would see their wealth grow over a short time period, while others would see it shrink. It might look like wealth is being transferred between individuals but, mechanistically in the model, these apparent transfers are the results of random fluctuations.

We wish to model a population of individuals whose wealths are connected through social and economic structures such as governments and markets. Such a population will likely exhibit resource allocation that is not captured by GBM alone. The mechanism we introduce to model these interactions creates systematic transfers of wealth between individuals. We stress that these are over and above the apparent transfers observed in a population whose wealths follow GBM.

Our reallocation mechanism has many possible correspondences to aspects of real economies. With $\tau>0$ it resembles a flat wealth tax with equal redistribution, while with $\tau<0$ it is akin to the collection of rents.

Governments collect taxes and spend revenues on welfare and public goods. Therefore, we might surmise that the net effect of the public sector is to reallocate from richer to poorer ($\tau>0$). However, it is not clear who benefits most from public spending. The richer might benefit disproportionately from a legal system to protect property rights; or from centrally-funded infrastructure, such as power and transport networks, which they use not only personally but also to further commercial interests.

Likewise, it is unclear who benefits most from the private sector. The flows of wealth to owners of assets which provide essential services -- such as houses, supermarkets,  and telephone networks -- look like negative reallocation ($\tau<0$). Everyone pays for the services at a fixed rate and the `rents' thus collected are distributed in proportion to wealth. On the other hand, the procurers of the services may derive a net benefit from them, and the provision of such services usually requires the payment of wages, \ie fixed-rate disbursements to many people.

The argument who benefits more cannot be settled by long lists of examples. It should be addressed empirically. We find a single time-varying value of $\tau$ -- an effective reallocation rate summarising the net action of all parts of the economy, both public and private -- which best describes observed wealth inequalities in the United States.

\section{Empirical study}
\subsection{Estimating $\mu$}
We estimate $\mu$ from historical private wealth data for the US \cite{PikettyZucman2014} by dividing the total private wealth by the total population size. In the analysed data ``private wealth [...] is the net wealth (assets minus liabilities) of households and non-profit institutions serving households. [...] Assets include all the non-financial assets -- land, buildings, machines, and so on -- and financial assets, including life insurance and pensions funds, over which ownership rights can be enforced and that provide economic benefits to their owners'' \cite[p.~1268]{PikettyZucman2014}.

Starting at $t_0=1917$ years AD, we assume the average historical private wealth follows $\exp\left[\mu\left(t-t_0\right)\right]$ and find $\mu=0.021\pm 0.001 \text{ year}^{-1}$ using a least-squares regression (\fref{mu_es}). For the timescale of interest, about $100$ years, the growth rate of the population average is about the growth rate of the expectation value, since the population is large enough (see Model section and also \cite{PetersKlein2013} and \cite{Bouchaud2015}).

\begin{figure}[!htb]
\centering
\includegraphics[width=0.5\textwidth] {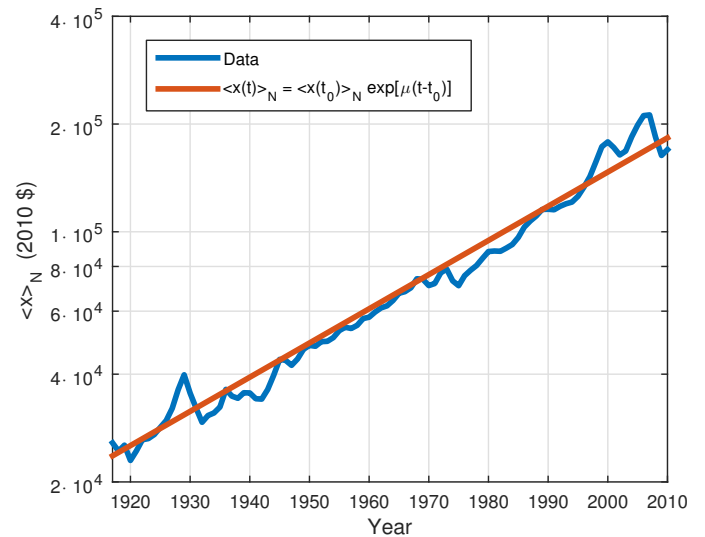}
\caption{Historical wealth per capita in the US (blue) and fitted exponential function with $\mu = 0.021 \text{ year}^{-1}$ (red). Data from \cite{PikettyZucman2014}.
\flabel{mu_es}}
\end{figure}

\subsection{Estimating $\sigma$}
For each year we estimate $\sigma(t)$ as the standard deviation of daily logarithmic changes, annualised by multiplying by $(250/\text{year})^{1/2}$, of the Dow Jones Industrial Average (data from Quandl database \cite{Quandl2016}). The values usually lie within the range $0.1-0.2 \text{ year}^{-\frac{1}{2}}$, with an average of $0.16$. Using a fixed $\sigma$ within the range $0.1-0.2$ has little effect on our results.

\subsection{Inferring $\tau$}
Our key interest is to fit a time series, $\tau\left(t\right)$, that reproduces the annually observed wealth shares in \cite{SaezZucman2014}. The wealth share $S_q$ is defined as the proportion of total wealth, $\sum_i^N x_i$, owned by the richest fraction $q$ of the population, \eg $S_{10\%}=80\%$ means that the richest 10\% of the population own $80\%$ of the total wealth.

For an empirical wealth share time series, $\Sdata_q\left(t\right)$, we proceed as follows.

\bi
\item[Step 1]
Initialize $N$ individual wealths, $\{x_i(t_0)\}$, as random variates of a lognormal distribution\footnote{The lognormal is a reasonable representation of observed distributions \cite{Sargan1957distribution}, and is generated by \eref{dx2} for $\tau=0$. Our results do not depend strongly on the initial distribution.} whose parameters are chosen to match $\Sdata_q(t_0)$.
\item[Step 2]
Propagate $\{x_i\left(t\right)\}$ according to \eref{dx2} over $\Dt=1$ year, using the value of $\tau$ that minimizes the difference between the wealth share in the modelled population, $\Smodel_q(t+\Dt, \tau)$, and $\Sdata_q(t+\Dt)$. We use the Nelder-Mead algorithm \cite{NelderMead1965}.
\item[Step 3]
Repeat Step 2 until the end of the time series in 2012.
\ei

We consider historical wealth shares of the richest $q=10\%$, $5\%$, $1\%$, $0.5\%$, $0.1\%$ and $0.01\%$ and obtain time series of fitted effective reallocation rates, $\tau_q\left(t\right)$, shown in \fref{tau}. For each value of $q$ we perform $10$ independent runs of the simulation for $N=10^6$ and average over the obtained results. Since in practice $dW$ is randomly chosen, each run of the simulation will result in slightly different $\tau_q\left(t\right)$ values. However, the differences between such calculations are negligible. We observe a short spin-up period of approximately 3 years, after which $\tau_q\left(t\right)$ is no longer sensitive to the initial distribution.

\begin{figure*}[!htb]
\centering
\includegraphics[width=1.0\textwidth] {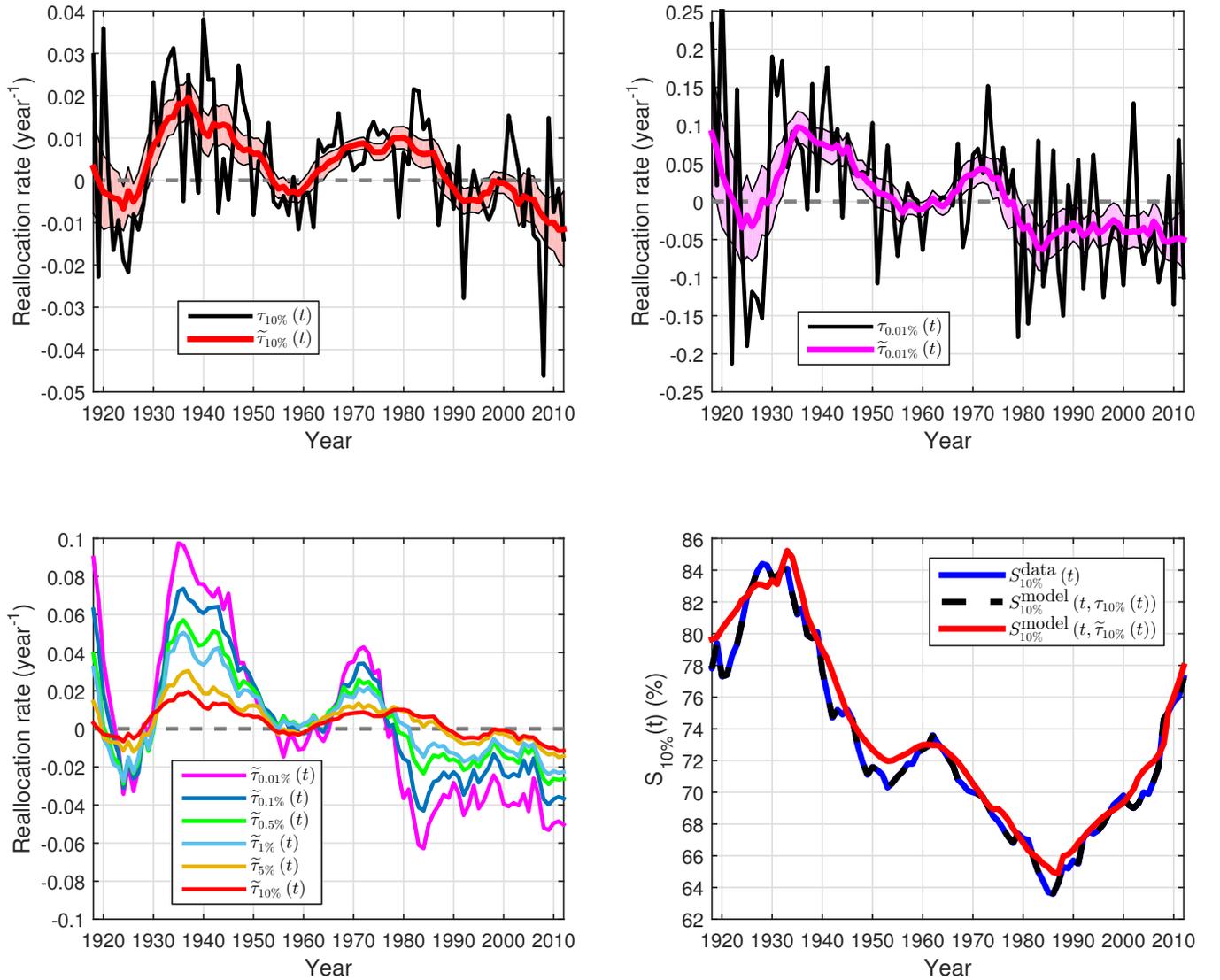}
\caption{Fitted effective reallocation rates using $\mu=0.021 \text{ year}^{-1}$ and Dow-Jones-derived $\sigma\left(t\right)$.
\underline{Top left:} $\tau_{10\%}\left(t\right)$ (black) and $\wtau_{10\%}\left(t\right)$ (red). \underline{Top right:} $\tau_{0.01\%}\left(t\right)$ (black) and $\wtau_{0.01\%}\left(t\right)$ (magenta). Translucent envelopes indicate one standard error in the moving averages. \underline{Bottom left:} $\wtau\left(t\right)_q$ reproducing the wealth shares listed in the legend. \underline{Bottom right:} $\Sdata_{10\%}$ (blue), $\Smodel_{10\%}$ based on the annual $\tau_{10\%}\left(t\right)$ (dashed black), based on the 10-year moving average $\wtau_{10\%}\left(t\right)$ (red).
\flabel{tau}}
\end{figure*}

\Fref{tau} shows large annual changes in $\tau_q\left(t\right)$. We are interested in longer-term changes in reallocation driven by structural economic and political changes. To elucidate these we take a central moving average, $\wtau_q\left(t\right)$, with a 10-year window truncated at the ends of the time series. This smoothing reveals a recognizable history of wealth reallocation, \eg the stock-market boom of the 1920s which benefited wealthy shareholders ($\wtau_q\left(t\right)<0$) and the social programs of the New Deal ($\wtau_q\left(t\right)>0$). 

The most important observation is that the effective reallocation rate has been negative since the mid-1980s. To test the robustness of this finding we repeated the analysis for other inequality measures: the Gini coefficient, the Kolmogorov-Smirnov statistic of the wealth cumulative distribution function, and the Lorenz curve based on $\Sdata_q\left(t\right)$. This makes little difference: the same behaviour of $\wtau\left(t\right)$, including the substantial negative values from the 1980s onward, is obtained for all inequality measures we analysed. 

To ensure that the smoothing does not introduce artificial biases, we reversed the procedure and used $\wtau_q\left(t\right)$ to propagate the initial lognormally distributed $\{x_i\left(t_0\right)\}$ and determine the wealth shares $\Smodel_q\left(t\right)$. The good agreement with $\Sdata_q\left(t\right)$ suggests that the smoothed $\wtau_q\left(t\right)$ is meaningful, see \fref{tau}.

\section{Discussion}
GBM, without reallocation, is a model of an economy without social structure. Wealth inequality in this model increases indefinitely. We considered this unrealistic, and expected the net effect of real social institutions to be reflected in positive $\tau$. Therefore, the negative reallocation rates we found came as a shock. We didn't imagine that explicitly reallocative structures such as taxation and welfare spending had been overridden to the point of reversal.

Having said that, with realistic parameters $\sigma$ and $\mu$, and non-negative $\tau$ everyone gets wealthier over time under the RGBM model. Therefore the recent observation that 
``the wealth owned by the bottom half of humanity has fallen by a trillion dollars in the past five years'' \cite[p.~2]{Hardoon2016} is consistent with negative $\tau$, not only in the US but also globally.


Studies of inequality often make the following assumptions that go under the headings of equilibrium, ergodicity, stationarity, or stability.

\bi
\item[A.] The system can equilibrate, \ie a stationary distribution exists to which the observed distribution converges in the long-time limit. For example, a recent study states: ``We impose assumptions [...] that guarantee the existence and uniqueness of a limit stationary distribution'' \cite[p.~130]{BenhabibBisinZhu2011}.
\item[B.] The system equilibrates quickly, \ie the observed distribution gets close to the stationary distribution after a time shorter than the timescale of observation. This assumption is often left unstated, but it is necessary for the stationary distribution to have practical relevance.
\ei

Such studies fit parameters of the stationary distribution which reproduce observed inequality. For example, a researcher would observe a wealth share and then find the reallocation rate, $\etau_q$, for which the model produces the same wealth share in the long-time limit.

We do not make the equilibrium assumptions. Fitting $\tau$ in RGBM allows the data to speak without constraint as to whether the assumptions are valid. We find them to be invalid because:

\bi
\item[A.] The system cannot equilibrate for $\tau \leq 0$. The model generates a non-stationary distribution and running it for longer produces greater inequality.
\item[B.] The system does not equilibrate quickly for realistic parameters. The shortest equilibration time we inferred from $\wtau_{10\%}$ was approximately 50 years when $\wtau_{10\%}$ was at its maximum of $0.02 \text{ year}^{-1}$. This is much longer than the observation time of one year.\footnote{The equilibration time was estimated as the exponential rate of convergence of the variance of the distribution of relative wealth $x_i/\ave{x}_N$ to its asymptotic value for prevailing levels of $\tau$ and $\sigma$.}
\ei

\Fref{asymptau} contrasts $\etau_{10\%}\left(t\right)$ as would be found in an equilibrium study with $\wtau_{10\%}\left(t\right)$ as found in this study. If equilibration were always possible and fast, then the two values would be identical within statistical uncertainties. They are not. 

\begin{figure*}[!htb]
\centering
\includegraphics[width=1\textwidth] {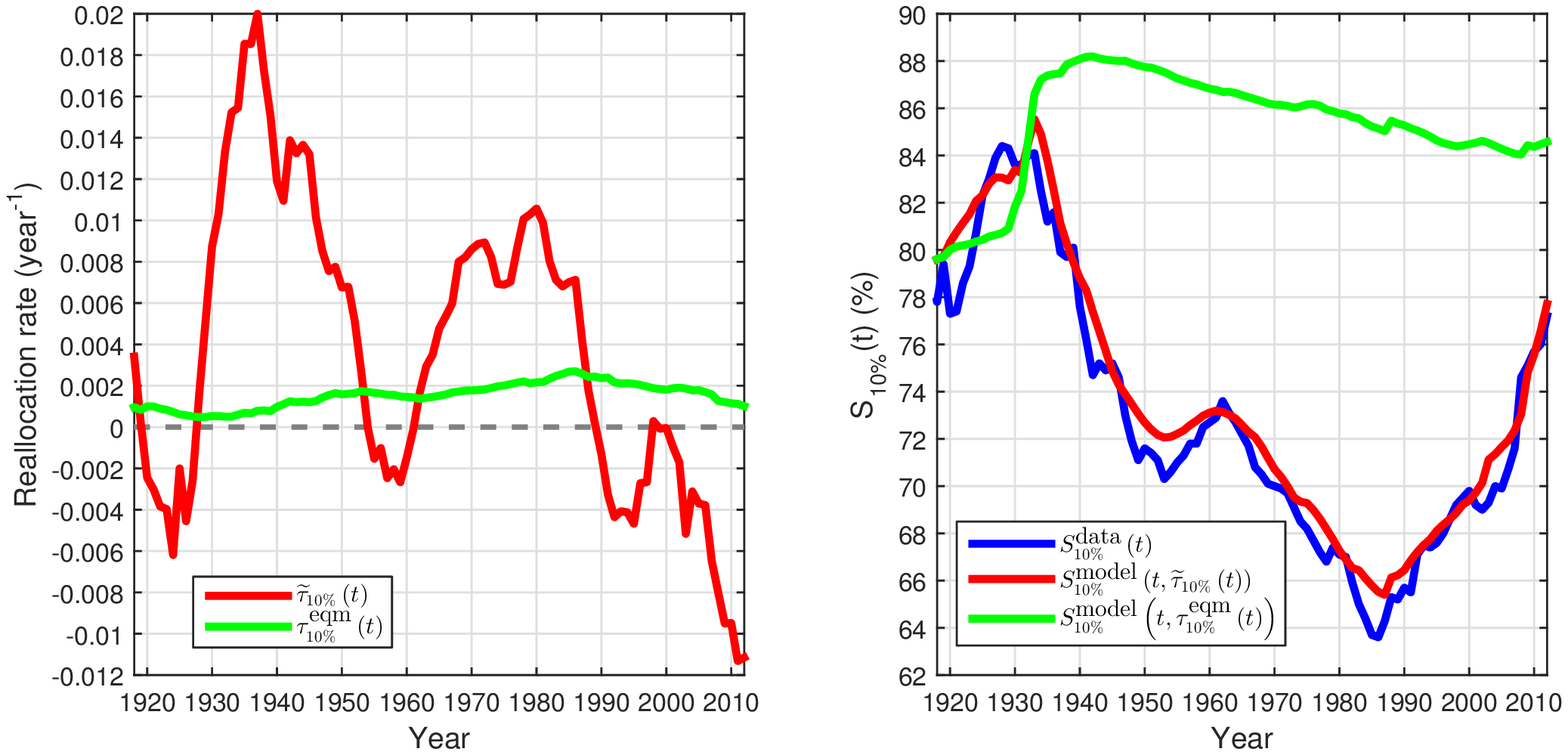}
\caption{Comparison of dynamic and equilibrium reallocation rates. \underline{Left:} $\wtau_{10\%}\left(t\right)$ (red, same as in the top left of \fref{tau}). $\etau_{10\%}\left(t\right)$ (green), defined such that $\lim_{t'\to\infty} \Smodel_{10\%}\left(t',\etau_{10\%}\left(t\right)\right)=\Sdata_{10\%}\left(t\right)$. It is impossible by design for this value to be negative. The significant difference between the red and green lines demonstrates that the equilibrium assumption is invalid for the problem under consideration. \underline{Right:} $\Sdata_{10\%}$ (blue), $\Smodel_{10\%}$ based on the 10-year moving average $\wtau_{10\%}\left(t\right)$ (red), based on $\etau_{10\%}\left(t\right)$ (green). The reallocation rates found under equilibrium assumptions generate model wealth shares which bear little relation to reality.
\flabel{asymptau}}
\end{figure*}

The equilibrium assumptions preclude what we find in this study, namely reallocation rates that correspond to wealth distributions which are either non-stationary ($\wtau_{q}\left(t\right)\leq 0$) or which equilibrate slowly (all observed $\wtau_{q}\left(t\right)>0$). Currently (\ie in 2012, the latest year of available data) the system is in a state best described in RGBM by $\tau < 0$. Each time we observe it, we see a snapshot of a distribution in the process of diverging. It is much like taking a photo of an explosion in space: it will show a fireball whose finite extent tells us nothing of the eventual distance between pieces of debris. Studies -- of both fireballs and inequality -- that assume equilibrium must be treated skeptically, as they are incapable of detecting the dramatic conditions one finds without this assumption.

\begin{acknowledgements}
YB acknowledges travel support and hospitality from London Mathematical Laboratory (LML). This research was partly funded by the LML Summer School 2015.
\end{acknowledgements}

\bibliographystyle{plain} 
\bibliography{inequality_bib2}

\begin{thebibliography}{10}
\expandafter\ifx\csname url\endcsname\relax
  \def\url#1{\texttt{#1}}\fi
\expandafter\ifx\csname urlprefix\endcsname\relax\def\urlprefix{URL }\fi
\expandafter\ifx\csname href\endcsname\relax
  \def\href#1#2{#2} \def\path#1{#1}\fi

\bibitem{PetersKlein2013}
O.~Peters, W.~Klein, Ergodicity breaking in geometric {B}rownian motion, Phys.
  Rev. Lett. 110~(10) (2013) 100603.
\newblock \href {http://dx.doi.org/10.1103/PhysRevLett.110.100603}
  {\path{doi:10.1103/PhysRevLett.110.100603}}.

\bibitem{BouchaudMezard2000}
J.-P. Bouchaud, M.~M\'ezard, Wealth condensation in a simple model of economy,
  Physica A: Statistical Mechanics and its Applications 282~(4) (2000)
  536--545.
\newblock \href {http://dx.doi.org/10.1016/S0378-4371(00)00205-3}
  {\path{doi:10.1016/S0378-4371(00)00205-3}}.

\bibitem{PetersAdamou2015a}
O.~Peters, A.~Adamou, \href{http://arxiv.org/abs/1506.03414}{The evolutionary
  advantage of cooperation}, ar{X}iv:1506.03414.
\newline\urlprefix\url{http://arxiv.org/abs/1506.03414}

\bibitem{PikettyZucman2014}
T.~Piketty, G.~Zucman, Capital is back: Wealth-income ratios in rich countries,
  1700-2010, The Quarterly Journal of Economics 129~(3) (2014) 1255--1310.
\newblock \href {http://dx.doi.org/10.1093/qje/qju018}
  {\path{doi:10.1093/qje/qju018}}.

\bibitem{Bouchaud2015}
J.-P. Bouchaud, \href{http://arXiv.org/abs/1508.00275}{On growth-optimal tax
  rates and the issue of wealth inequalities}, ar{X}iv:1508.00275.
\newline\urlprefix\url{http://arXiv.org/abs/1508.00275}

\bibitem{Quandl2016}
{D}ow {J}ones {I}ndustrial {A}verage,
  \url{https://www.quandl.com/data/BCB/UDJIAD1-Dow-Jones-Industrial-Average},
  accessed: 19/04/2016.

\bibitem{SaezZucman2014}
E.~Saez, G.~Zucman, Wealth inequality in the {U}nited {S}tates since 1913:
  Evidence from capitalized income tax data, Tech. rep., {National Bureau of
  Economic Research} (2014).
\newblock \href {http://dx.doi.org/10.3386/w20625} {\path{doi:10.3386/w20625}}.

\bibitem{Sargan1957distribution}
J.~D. Sargan, The distribution of wealth, Econometrica, Journal of the
  Econometric Society (1957) 568--590\href {http://dx.doi.org/10.2307/1905384}
  {\path{doi:10.2307/1905384}}.

\bibitem{NelderMead1965}
J.~A. Nelder, R.~Mead, A simplex method for function minimization, The Computer
  Journal 7~(4) (1965) 308--313.
\newblock \href {http://dx.doi.org/10.1093/comjnl/7.4.308}
  {\path{doi:10.1093/comjnl/7.4.308}}.

\bibitem{Hardoon2016}
D.~Hardoon, R.~Fuentes-Nieva, S.~Ayele, An economy for the 1\%: How privilege
  and power in the economy drive extreme inequality and how this can be
  stopped, Tech. rep., Oxfam International (2016).

\bibitem{BenhabibBisinZhu2011}
J.~Benhabib, A.~Bisin, S.~Zhu, The distribution of wealth and fiscal policy in
  economies with finitely lived agents, Econometrica 79~(1) (2011) 123--157.
\newblock \href {http://dx.doi.org/10.3982/ECTA8416}
  {\path{doi:10.3982/ECTA8416}}.

\end{thebibliography}

\end{document}